\documentclass[twocolumn]{aastex63}

\newcommand\be{\begin{eqnarray}}
\newcommand\ee{\end{eqnarray}}
\usepackage{amsmath}
\submitjournal{ApJL}
\shorttitle{Effects of spin on growth of early SMBHs}
\shortauthors{Zhang, Lu \& Fang}
\begin{document}
\title{Effects of spin on constraining the seeds and growth of $\gtrsim 10^9M_\odot$ supermassive black holes in $z>6.5$ Quasars}
\correspondingauthor{Xiaoxia Zhang}
\email{zhangxx@xmu.edu.cn; luyj@nao.cas.cn; fangt@xmu.edu.cn}
\author[0000-0003-4832-9422]{Xiaoxia Zhang}
\affiliation{Department of Astronomy, Xiamen University, Xiamen, Fujian 361005, China}

\author[0000-0002-1310-4664]{Youjun Lu}
\affiliation{National Astronomical Observatories, Chinese Academy of Sciences,  
20A Datun Road, Beijing 100101, China }
\affiliation{School of Astronomy and Space Sciences, University of Chinese Academy of Sciences, No. 19A Yuquan Road, Beijing 100049, China}
\author[0000-0002-2853-3808]{Taotao Fang}
\affiliation{Department of Astronomy, Xiamen University,  Xiamen, Fujian 361005, China}

\begin{abstract}
The existence of $\gtrsim10^9M_\odot$ supermassive black holes (SMBHs) at redshift $z>6$ raises the problem of how such SMBHs can grow up within the cosmic time ($<1$\,Gyr) from small seed BHs. In this Letter, we use the observations of $14$ quasars at $z>6.5$ with mass estimates to constrain their seeds and early growth, by self-consistently considering the spin evolution and the possibility of super-Eddington accretion. We find that spin plays an important role in the growth of early SMBHs, and the constraints on seed mass and super-Eddington accretion fraction strongly depend on the assumed accretion history. If the accretion is coherent with single (or a small number of) episode(s), leading to high spins for the majority of accretion time, then the SMBH growth is relatively slow; and if the accretion is chaotic with many episodes and in each episode the total accreted mass is much less than the SMBH mass, leading to moderate/low spins, then the growth is relatively fast. The constraints on the seed mass and super-Eddington accretion fraction are degenerate. A significant fraction ($\gtrsim0.1\%-1\%$ in linear scale but $\sim 3-4$ dex in logarithmic scale for $10^3-10^4 M_\odot$ seeds) of super-Eddington accretion is required if the seed mass is not $\gg10^{5}M_\odot$, and the requirements of high seed mass and/or super-Eddington accretion fraction are moderately relaxed if the accretion is chaotic.
\end{abstract}


\section{Introduction} 
\label{sec:intro}

Observations of high-redshift quasars suggest that $\gtrsim10^9 M_\odot$ supermassive black holes (SMBHs) are already in place at $z\gtrsim7$ \citep[e.g., see][for a review]{2019arXiv191105791I}, which raises the question whether the time is sufficient for the growth of such SMBHs \citep[e.g.,][]{2004ApJ...614L..25Y}. Theoretical studies and simulations indicate that the death of first (Population III) stars at $z\sim20-30$ results in BHs with mass $\sim10^2 M_\odot$ \citep[e.g.,][]{2014ApJ...781...60H}. To grow those light seeds to $\gtrsim10^9 M_\odot$ within hundreds of millions of years, a period of super-Eddington accretion must be invoked \citep[e.g., see][]{2012MNRAS.424.1461L, 2014ApJ...784L..38M, 2015ApJ...804..148V}. Alternatively, direct collapse of gas clouds, leading to heavier seeds of $\sim10^5 M_\odot$ \citep[e.g.,][]{2003ApJ...596...34B, 2005ApJ...633..624V}, may alleviate the need for super-Eddington accretion. 

The radiative efficiency ($\eta$) is also an important parameter for the SMBH mass growth as it directly determines the $e$-folding timescale (or the Salpeter timescale $\tau_{\rm Sal} = 4.5\times 10^8  \frac{\eta}{1-\eta}$\,yr). 
This timescale may vary by a factor of $\sim 7$ for spin in the range $0-0.998$ and correspondingly $\eta \sim 0.057-0.31$ at least in the thin-disk accretion regime \citep[e.g.,][]{1970Natur.226...64B, 1974ApJ...191..507T}, which suggests the spin evolution should be considered when addressing the growth problem of early SMBHs. However, $\eta$ is commonly set as the canonical value of $0.1$ \citep[e.g.,][]{2002MNRAS.335..965Y} in most previous works. The ignorance of the SMBH spin evolution and the effect of radiative efficiency may lead to inaccurate constraints on the seed mass and the significance of the super-Eddington accretion.  

In this Letter, we make use of a sample of $z>6.5$ quasars to constrain the seeds and growth of early $\gtrsim 10^9M_\odot$ SMBHs, taking into account self-consistently both the spin evolution and the possibility of super-Eddington accretion. In particular, we demonstrate that a degeneracy exists between the constraints on the seed mass and contribution fraction of super-Eddington accretion, and different accretion histories may result in significantly different constraints. The Letter is organized as follows. We present possible accretion histories and SMBH evolution in Section~\ref{sec:history}, followed by a description of the data and statistical method in Section~\ref{sec:data}. Results and discussion are given in Section~\ref{sec:result}, and conclusions are summarized in Section~\ref{sec:cons}. Throughout the Letter, we adopt a flat cosmology with $H_0=70 \ {\rm km\,s^{-1}\,Mpc^{-1}}$, $\Omega_M=0.27$, and $\Omega_\Lambda=0.73$. 

\section{SMBH accretion histories and growth}
\label{sec:history}

We adopt several toy models to describe the accretion histories of these early SMBHs, involving both super-Eddington and sub-Eddington accretion, which may be typically experienced by them. We define $f_{\rm sup}$ as the fraction of mass growth contributed by the super-Eddington accretion, with $f_{\rm sup}=0$ representing the cases without super-Eddington accretion and $f_{\rm sup}=1$ representing the cases with super-Eddington accretion only. For simplicity, we adopt a constant accretion rate of $\dot{m} \equiv \dot{M}/\dot{M}_{\rm Edd}=10$, where $\dot{M}_{\rm Edd}=16 L_{\rm Edd}/c^2$ with $L_{\rm Edd}$ the Eddington luminosity. 
The duration of an accretion episode is controlled by the disk mass and accretion rate (or Eddington ratio), and can be approximated as $\Delta t_{\rm epi}\sim M_{\rm disk}/\dot{M}$. After that, a new accretion episode instantaneously starts, which means the duty cycle is assumed to be unity in this work.
Therefore, the constraints on the seed mass and $f_{\rm sup}$ obtained in Section~\ref{sec:result} could be lower limits if the real duty cycle is substantially smaller than $1$. To figure out how the seed mass correlates with $f_{\rm sup}$ under different accretion and thus different spin evolution histories, we consider the following three toy models.
\begin{itemize}
\item[1.]  Model A: The SMBH accretes continuously and coherently. The accretion rate is super-Eddington until the mass reaches $M_{\rm crit}=M_{\bullet, \rm s}+f_{\rm sup} (M_{\bullet, \rm f}-M_{\bullet, \rm s})$ with $M_{\bullet, \rm s}$ the seed mass and $M_{\bullet, \rm f}$ the final mass.  After that, the accretion rate drops to sub-Eddington and the thin-disk criterion is satisfied. 
\item [2.] Model B: The SMBH experiences an initial phase of supercritical accretion followed by multi-episode chaotic thin-disk accretion phase with a random disk orientation in each episode. Similar to Model A, the SMBH maintains the super-Eddington rate before $M_{\bullet}= M_{\rm crit}$. After that, the accretion rate transits to sub-Eddington, and the disk mass in each chaotic episode scales with the SMBH mass \citep[see also][]{2019ApJ...873..101Z}, i.e.,
\be
M_{\rm disk}=b M_\bullet \left( \frac{M_\bullet}{10^8 M_\odot}\right)^\gamma,
\label{eq:Mdisk}
\ee
For demonstration purposes, we fix $b=0.01$ and $\gamma=0.5$, according to the constraints from \citet[][]{2019ApJ...873..101Z}. The set of $\gamma=0.5$ can avoid extreme spins at early time, and it makes a distinct spin evolution from that of Model A.\footnote{We note here that a smaller $\gamma$ (e.g., $\gamma=0$) means a larger $M_{\rm disk}$ at early time and a smaller $M_{\rm disk}$ at late time, and will give rise to higher spins before $M_\bullet=10^8 M_\odot$ and to lower spins after that, indicating slow mass growth at early time and faster growth at late time. By the fact that earlier growth is more efficient in saving time, a smaller $\gamma$ will shift the $M_{\bullet, \rm s}$ vs. $f_{\rm sup}$ contours toward the upper right. A smaller $b$ will result in lower spins on average and thus more efficient mass growth, leading to a requirement for lighter seeds and/or a lower contribution from super-Eddington accretion.} 
\item [3.] Model C: The accretion history is composed of multiple accretion episodes and in each episode the accretion rate declines from super-Eddington to sub-Eddington. Typical timescales of SMBH mergers revealed by cosmological simulations indicate that the number of the merger events that a BH at $z\sim7$ has experienced is $\lesssim10$ \citep[e.g.,][]{2015MNRAS.449...49R}. We therefore take $N=10$  as the total number of episodes, and the logarithmic mass increase in each episode is assumed to be the same, i.e., $\Delta \log M_\bullet=(M_{\bullet, \rm f}-M_{\bullet, \rm s})/N$, and in each $i$th episode, $f_{\rm sup}$ also defines an initial part of super-Eddington accretion and the rest of sub-Eddington accretion as that in Model A.
\end{itemize}

The mass and spin evolution for SMBHs under different accretion modes has been investigated in our previous works \citep[e.g.,][]{2019ApJ...873..101Z, 2019ApJ...877..143Z, 2020ApJ...896...87Z}, and we briefly summarize it as follows.

For a BH accreting under the models described above, the evolution of its mass $M_\bullet$ and spin vector ${\bf J}_\bullet$ can be described by
\be
\frac{d M_\bullet}{d t} &=& f_{\rm Edd} \frac{1-\eta}{\eta} \frac{M_\bullet}{\tau_{\rm Edd}},
\label{eq:dmdt}
\ee
\be
\frac{d{\bf J}_\bullet}{d t} &=& \dot{M} \frac{G M_\bullet}{c}\Phi(R_{\rm in})\hat{\bf l}+\frac{4\pi G}{c^2} \int_{\rm disk} \frac{{\bf L}\times {\bf J}_\bullet}{R^2} d R, 
\label{eq:djdt}
\ee
(for each accretion episode), where $f_{\rm Edd}$ is the Eddington ratio and $\tau_{\rm Edd} = 4.5\times 10^8\ {\rm yr}$ is the Eddington timescale.
The first term on the right side of Equation~\eqref{eq:djdt} denotes the momentum injection at the inner disk boundary $R_{\rm in}$, where $\dot{M}$ is the accretion rate, $G$ is the gravitational constant, $c$ is the speed of light, $\Phi$ is the specific angular momentum, and $\hat{\bf l}$ is a unit vector paralleled with the angular momentum at $R_{\rm in}$; the second term describes the frame-dragging torque due to the misalignment between the angular momenta of the disk and SMBH, with ${\bf L}$ the angular momentum of the disk per unit area.

For super-Eddington accretion, the accretion disk is thick in geometry and the inner disk boundary is slightly different from that of the thin disk. Ignoring the initial short period for the alignment, the SMBH is spun up until it reaches the canonical value of $0.998$ \citep{1974ApJ...191..507T}. Equation~\eqref{eq:djdt} is then reduced to 
\be
\frac{da}{dt}=[\Phi (R_{\rm in})-2a(1-\eta)] \frac{f_{\rm Edd}}{\eta \ t_{\rm Edd}},
\label{eq:dadt}
\ee
where $a$ is the dimensionless spin parameter and $|a| \equiv c|{\bf J}_\bullet|/(GM^2_\bullet)$. 

For sub-Eddington accretion, the standard thin-disk model is adopted, and the disk warping due to the Bardeen-Petterson effect is considered until the BH spin aligns with the angular momentum of the outer disk. Then Equation~\eqref{eq:dadt} is solved \citep[see][for details]{2019ApJ...873..101Z}.

\section{Observations versus Model Objects}
\label{sec:data}

\begin{deluxetable}{ccccc}
\tablenum{1}
\tablecaption{  Quasars at $z>6.5$ with both mass and Eddington ratio estimations. \label{tab:1}}
\tablehead{
\colhead{Object  Name} 	& \colhead{$z$}   & \colhead{$M_\bullet (10^9 M_\odot)$} 	& \colhead{$f_{\rm Edd}$}	& \colhead{References} } 
\startdata
J1342+0928 	& 7.541 & $0.91^{+0.14}_{-0.13}$ & $1.1\pm0.2$            & 1 \\
J1007+2115	& 7.515	& $1.5\pm0.2$            & $1.06\pm0.2$	          & 2 \\
J1120+0641	& 7.087	& $2.47^{+0.62}_{-0.67}$ & $0.57^{+0.16}_{-0.27}$ & 3 \\
J1243+0100	& 7.07	& $0.33\pm0.20$ 	     & $0.34\pm0.2$	          & 4 \\
J0038-1527	& 7.021	& $1.33\pm0.25$	         & $1.25\pm0.19$	      &	5 \\
J2348-3054	& 6.902	& $1.98^{+0.57}_{-0.84}$ & $0.17^{+0.92}_{-0.88}$ & 3 \\
J0109-3047	& 6.763	& $1.33^{+0.38}_{-0.62}$ & $0.29^{+0.88}_{-2.59}$ & 3 \\
J1205-0000	& 6.73	& $4.7^{+1.2}_{-3.9}$ 	 & $0.06^{+0.32}_{-0.58}$ & 3 \\
J338+29		& 6.658	& $3.7^{+1.3}_{-1.0}$	 & $0.13^{+0.05}_{-0.04}$ & 6 \\
J0305-3150	& 6.610	& $0.90^{+0.29}_{-0.27}$ & $0.64^{+2.20}_{-3.42}$ & 3 \\
J323+12		& 6.592	& $1.39^{+0.32}_{-0.51}$ & $0.44^{+1.09}_{-3.19}$ & 3 \\
J231-20		& 6.587	& $3.50^{+0.44}_{-2.24}$ & $0.48^{+0.11}_{-0.39}$ & 3 \\
J036+03		& 6.527	& $1.9^{+1.1}_{-0.8}$	 & $0.96^{+0.70}_{-0.35}$ & 6 \\
J167-13		& 6.508	& $0.49\pm0.20$	         &  $1.2\pm0.5$		      & 6 \\
\enddata
\tablecomments{Columns from left to right represent (1) object name, (2) redshift, (3) BH mass, (4) Eddington ratio, and (5) the references that provide the mass and Eddington ratio measurements. References: 1=\citet{2020ApJ...898..105O}, 2=\citet{2020ApJ...897L..14Y}, 3=\citet{2017ApJ...849...91M}, 4=\citet{2019ApJ...872L...2M}, 5=\citet{2018ApJ...869L...9W}, 6=\citet{2015ApJ...801L..11V}. }
\end{deluxetable}

We consider those quasars at $z>6.5$ that have both SMBH mass and Eddington ratio estimations as listed in Table~\ref{tab:1}. Those $14$ SMBHs weigh $\sim(0.3-5) \times 10^9 M_\odot$, and their Eddington ratios cover a broad range from $\lesssim0.1$ to $\gtrsim 1$. \citet{2020MNRAS.496..888N} also considered quasars at this redshift range, while they only took the mass information of eight of them with smaller errors, and their aim was to demonstrate the effect of the Hubble parameter on the seeding machanism of BHs.Below we generate mock samples to match with the observational data. 

For given accretion models and parameters $(f_{\rm sup},\ M_{\bullet, \rm s})$, we consider a BH population seeded at $z=25$, and their initial spins are randomly drawn from a uniform distribution between $0$ and $1$. Their final masses are confined in the range between $10^9$ and $10^{10} M_\odot$, where we evenly take $20$ values, in logarithmic space, as the final masses. For each final mass, we simulate $100$ BHs with Monte Carlo procedure, resulting in different accretion histories in terms of disk orientation and Eddington ratio in each episode. The Eddington ratio of a thin-disk episode is drawn from a Gaussian distribution with the mean of $0.68$ and standard deviation of $0.65$, which is obtained by fitting to a large mock sample generated from the observed mean and standard deviation of $f_{\rm Edd}$ in Table~\ref{tab:1}. We then set lower and upper boundaries of $0.1$ and $1$ to $f_{\rm Edd}$ in order to be roughly consistent with the observations. For those $2000$ BHs, the mass and spin evolution can be obtained by solving Equations~\eqref{eq:dmdt} and \eqref{eq:djdt}, and the bolometric luminosity ($L=\eta \dot{M}c^2$) evolution can be traced since the radiative efficiency $\eta$ is a function of spin. That means, for each BH, we have information about the mass, spin, and luminosity at different redshifts.

For the $i$the observed source (at $z_i$) in Table~\ref{tab:1}, we select mock samples at redshift $z_i$ from the $2000$ simulated BHs. Those mock samples are required to have luminosity within the observed uncertainties. The median masses ($M_{i, \rm the}$) of those mock samples are treated as the theoretical expectation of the model, and the $1\sigma$ uncertainty ($\sigma_{i, \rm the}$) is obtained through the $16th$ and $84th$ percentiles. Then the masses of the mock samples are compared with the observational ones according to the least $\chi^2$ estimator, i.e.,  
\be
\chi^2=\sum^N_{i=1} \frac{(\log M_{i, \rm the}-\log M_{i, \rm obs})^2}{\sigma^2_{i, \rm obs}+\sigma^2_{i, \rm the}},
\label{eq:chi2}
\ee
where $N$ is the total number of the observed source, $M_{i, \rm obs}$ is the mass, and $\sigma_{i, \rm obs}$ is the mass error (in unit of dex) of the $i$th source in Table~\ref{tab:1}. The errors $\sigma^2_{i, \rm the}$ of the model are considered because a mock object cannot be directly assigned to an observational one, while they can be paired statistically with given model mass errors of the mock object.

We divide the parameter space $(f_{\rm sup},\ M_{\bullet, \rm s})$ into grids, with $\Delta \log M_{\bullet, \rm s}=0.1$, and $\Delta \log f_{\rm sup}=0.1$ or $\Delta f_{\rm sup}=0.01$ depending on the whether $f_{\rm sup}$ is set as logarithmic or linear scale in the plot. For each grid with given $f_{\rm sup}$ and  $M_{\bullet, \rm s}$, we calculate $\chi^2$ values according to Equation~\eqref{eq:chi2}, and the minimum $\chi^2$ value defines the best-fit parameters, corresponding to a reduced value of $\chi_{\rm min}^2/\nu$, with $\nu$ the number of degrees of freedom. The cases here have $14$ data points and $2$ parameters ($\nu=12$); 
the $1\sigma$, $2\sigma$, and $3\sigma$ confidence levels correspond to $\Delta\chi^2$ values of $13.7$, $21.0$, and $29.8$, respectively, with respect to $\chi_{\rm min}^2$.

The above settings produce a population of BHs with final masses evenly distributed between $10^9$ and $10^{10} M_\odot$. We test an alternative choice that the simulated BHs have final masses in the same range but follow the mass distribution of active BHs at $z=6$ \citep[e.g.,][]{2010AJ....140..546W}. We find that our results in Section~\ref{sec:result} are robust against different assumptions on the final mass distribution of the simulated BHs.

\begin{figure*}
\centering
\includegraphics[width=\textwidth]{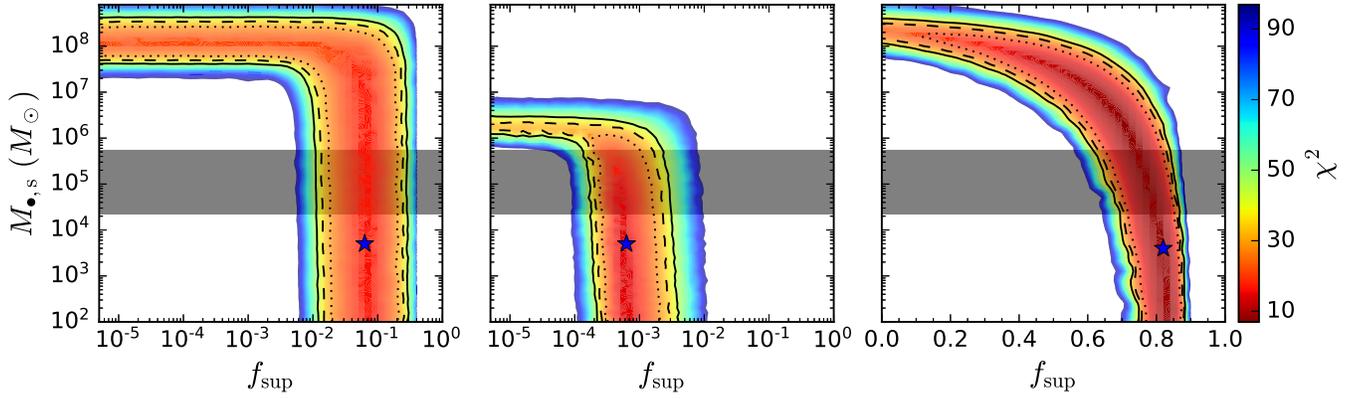}
\caption{Constraints on the mass of seed BH and the fraction of mass growth contributed by the super-Eddington accretion. Left, middle, and right panels show the results obtained from the data listed in Table~\ref{tab:1} for $14$ quasars at $z>6.5$ by assuming the BH growth Models A, B, and C, respectively. Here, Models A, B, and C represent those high-redshift quasars grew up via a single-epoch coherent accretion (an initial super-Eddington accretion phase followed by a thin-disk accretion phase with constant disk orientation), an initial super-Eddington accretion phase followed by multiple-episode chaotic thin-disk accretion, and multiple-episode accretion with the accretion rate in each episode declining from super- to sub-Eddington, respectively (see details in Section~\ref{sec:history}). The colors represent $\chi^2$ values estimated from Eq.~\eqref{eq:chi2}, and the dotted, dashed, and solid contours show $1\sigma$, $2\sigma$, and $3\sigma$ confidence levels, respectively. The blue star marks the location of the minimum $\chi^2$. The gray area is to highlight the seed masses of the samples in Table~\ref{tab:1} estimated by assuming the canonical radiative efficiency of $\eta=0.1$ and ignoring the contribution from super-Eddington accretion. For clarity, we adopt a linear scale for the x-axis of the right panel, but logarithmic scale for the x-axis of the left and middle panels. 
}
\label{fig-1}
\end{figure*} 

\section{Results and discussions}
\label{sec:result}

Figure~\ref{fig-1} shows constraints on the seed mass and contribution fraction of super-Eddington accretion for the three assumed accretion histories. As seen from this Figure, different accretion histories can lead to quite different constraints. A common trend is that a lighter seed requires a larger contribution fraction of the super-Eddington accretion, as expected. 

\begin{figure*}
\centering
\includegraphics[width=\textwidth]{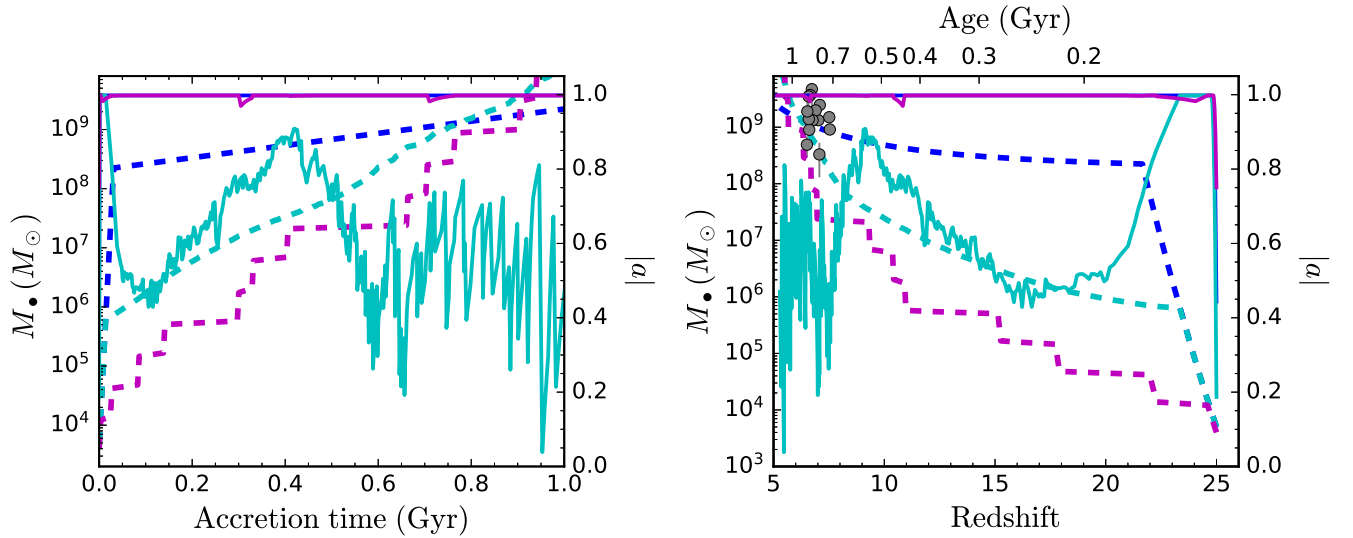}
\caption{Mass and spin magnitude evolution of several  SMBHs generated by assuming the growth Models A, B, and C, respectively. The left panel shows the mass (left y-axis; dashed lines) and spin magnitude (right y-axis; solid lines) evolution as a function of the accretion time. The right panel is similar but plot as a function of redshift/cosmic age. Blue, cyan, and magenta curves represent the results generated from the accretion Models A, B, and C, respectively, with the best-fit model parameters (blue stars in Fig.~\ref{fig-1}). The initial spins for these example objects are randomly generated. In the right panel, the filled circles represent the masses of the samples listed in Table~\ref{tab:1}.
}
\label{fig-2}
\end{figure*}

Figure~\ref{fig-2} shows the mass growth and spin magnitude evolution of three example SMBHs as a function of the accretion time/redshift/cosmic age. It illustrates significant differences of mass growth and spin evolution tracks of early SMBHs with different accretion histories as described by Model A, B, and C, respectively.

For Model A with continuously coherent accretion (left panel of Fig.~\ref{fig-1}), the SMBH is quickly spun up to the maximum value of $0.998$ and the spin maintains afterward (solid blue lines in Fig.~\ref{fig-2}). That means for most of its lifetime, the SMBH radiates with an efficiency of $\sim0.31$ \citep{1974ApJ...191..507T}, and the mass growth is quite inefficient. Therefore, growing these early $\sim 10^9M_\odot$ SMBHs requires an extremely large seed mass of about $10^8 M_\odot$ with negligible supercritical accretion or a relatively large ($\sim10\%$) contribution from super-Eddington accretion for lighter seeds. For comparison, the gray area marks the permitted seed masses of the observed sample in Table~\ref{tab:1} if they never underwent super-Eddington accretion, and the Eddington ratio and radiative efficiency are both constant with $f_{\rm Edd}=0.68$ and $\eta=0.1$ \citep[see also][]{2002MNRAS.335..965Y}. 

For Model B with continuous accretion followed by periods of chaotic thin-disk accretion, an extremely large seed is not necessarily required. However, if super-Eddington accretion is negligible, i.e., $f_{\rm sup} < 10^{-4}$, it still requires a seed mass of $\sim 10^6 M_\odot$. Nevertheless, a contribution fraction of $10^{-3}$ is sufficient for a seed of $10^2 M_\odot$ growing to $\sim 10^9 M_\odot$ by $z\sim7$. The chaotic phase causes the spin to oscillate over a broad range from about $0.2$ to $0.9$, and for most of the time, the spin has an intermediate value of $\sim 0.5-0.8$ (solid cyan curves in Fig.~\ref{fig-2}). This is the reason that the BH growth is more efficient than the case of Model A. For chaotic thin-disk accretion, the spin evolution strongly depends on the disk mass in each episode. For the power-law dependence form of $M_{\rm disk}$, as mentioned in Section~\ref{sec:history}, different choices of $\gamma$ in Equation~\eqref{eq:Mdisk} will result in different constraints, i.e., a smaller $\gamma$ requires a larger seed and/or a higher fraction of super-Eddington accretion. In addition, $f_{\rm Edd}$ will affect the growth timescale of SMBHs. However, it has little impact on the spin evolution as a function of mass. Therefore, the constraints mainly rely on the mean of $f_{\rm Edd}$ over the episodes, and a smaller $f_{\rm Edd}$ will require a larger seed and/or a larger $f_{\rm sup}$. We will further discuss the effect of the accretion rate below.

For Model C with multiple accretion episodes and in each episode super-Eddington accretion contributing a fraction $f_{\rm sup}$ to the mass growth in that episode, it requires either an extremely large seed mass ($\sim10^8 M_\odot$) without super-Eddington accretion or a contribution fraction of $80\%$ by super-Eddington accretion for a seed of $10^2 M_\odot$. For the total episodes of $10$ and the same mass increase in units of dex assumed here, the disk mass in each episode is $\gtrsim10\%$ of the BH mass, and disk angular momentum dominates over the SMBH spin \citep[e.g.,][]{2019ApJ...873..101Z}. In this case, the BH spin is always realigned to the disk momentum and the spin increases efficiently. Therefore, the BH stays at the maximum spin value for most of its lifetime (solid magenta lines in Fig.~\ref{fig-2}). Although with similar spin evolution as Model A, the mass growth is quite different because the sub-Eddington accretion is distributed over each episode. Since $d\ln M_\bullet \propto \tau_{\rm Sal}$, it takes the same time to grow a $10^2 M_\odot$ BH, for example, to $10^3 M_\odot$ and a $10^8 M_\odot$ BH to $10^9 M_\odot$, which means sub-Eddington accretion at early epoch consumes most of the time that allows an SMBH at $z\sim7$ to grow up (see Fig.~\ref{fig-2}), as the cosmic age at that time is only about $4 \tau_{\rm Sal}$ for $\eta=0.31$. Therefore, the super-Eddington accretion should play an overwhelming role in the growth of early SMBHs if the accretion history is more or less described by the Model C, though the time it takes is only a small fraction of the total accretion time (see the dashed magenta lines in Fig~\ref{fig-2}).

In our models, we assume a constant accretion rate for the super-Eddington phase and within each thin-disk episode. This is done to make the comparison among different models more straightforward. A more realistic case could be a time-evolving accretion rate. We argue that our results are not sensitive to this choice. For super-Eddington accretion, what matters is whether the SMBH can grow within a short time, compared to the sub-Eddington case. This can be achieved if the time-averaged accretion rate is larger than several times of the sub-Eddington rate. For thin-disk accretion, since $f_{\rm Edd}$ mainly affects the growth rate instead of the spin evolution against the SMBH mass, it is still the time-averaged $f_{\rm Edd}$ that determines the location of the contours shown in Fig.~\ref{fig-1}. Although the Gaussian distribution of $f_{\rm Edd}$ is obtained through fitting the data, it still suffers from small-sample statistics. The mean of the Gaussian function determines the center of the contours, and if the mean is larger, then the center shifts to smaller $M_{\rm s}$ and $f_{\rm sup}$; otherwise, to the opposite. The deviation of the Gaussian function exhibits some but not a large effect on the area of the contours for Models A and C, while for Model B, the contour area is mostly determined by `random' oscillations of the BH spin in the chaotic episodes. 

The samples in Table~\ref{tab:1} have Eddington ratios spanning a broad range and with large uncertainties. Some of them may be accreting at a super-Eddington rate. Our models with the current settings may not reproduce all SMBHs with the same masses and Eddington ratios as the samples. However, our main goal is not to simultaneously fit the masses and Eddington ratios of those high-$z$ objects, which can always be done by specifically setting the accretion rate in our models. Instead, we aim to demonstrate that different accretion histories may in general result in different constraints on the seed mass and contribution fraction by super-Eddington accretion. We therefore adopt the $f_{\rm Edd}$ distribution and an Eddington-limited boundary for thin-disk accretion for reference.

We do not include coalescence of BHs in our models. For two SMBHs of comparable masses, their mergers will result in a spin value of $\sim 0.7-0.9$ \citep[e.g.,][]{2010RvMP...82.3069C, 2010CQGra..27k4006L}, and will leave little long-term effect on the spin evolution \citep[e.g.,][]{2008MNRAS.385.1621K}. This case can be simply considered in our models by injecting the merger events into the accretion histories \citep[see][]{2008ApJ...689..732Y}, leading to a flip of spin and mass. This spin flip will quickly be washed out by the accretion of gas since the typical timescale of spin change is comparable to $\tau_{\rm Sal}$. What matters is whether the merger is efficient in growing mass compared with gas accretion. If $\tau_{\rm Sal}$ is larger than the merger timescale, then merger is more efficient in mass growth, giving rise to a shift of the contours to lower left.

\section{Conclusions}
\label{sec:cons}
By utilizing a sample of $z>6.5$ quasars and taking into account self-consistently the spin evolution and possibility of supercritical accretion, we obtain constraints on the seed BH mass and fraction contributed by super-Eddington accretion to the growth of early $\gtrsim 10^9M_\odot$ SMBHs. We find that the BH spin has important effects on these constraints. For accretion histories dominated by a coherent infall of gas clouds (e.g., with a small number of episodes), the spin keeps high values and the mass growth is inefficient, leading to a requirement for high-mass seeds of up to $\sim 10^8 M_\odot$ if they are without super-Eddington contribution. For accretion histories dominated by small episodes with random directions of the infalling clouds, the spin will oscillate around an intermediate value and the mass growth is faster, alleviating the requirements for extremely massive seeds if super-Eddington accretion is negligible. Current seeding mechanisms proposing a seed mass not larger than $10^5-10^6M_\odot$ call for a period of super-Eddington accretion, which contributes at least a fraction of $\gtrsim 0.1\%-1\%$ in linear scale (but $3-4$ dex in logarithmic mass scale  for $10^3-10^4 M_\odot$ seeds) to the mass growth.

\acknowledgements
We thank the anonymous referee for his/her helpful suggestions and comments. This work is supported by the National Key Program for Science and Technology Research and Development under Nos. 2017YFA0402600 and 2016YFA0400704; the National Natural Science Foundation of China under Nos. 11525312, 11890692, 11873056, 11690024, 11991052, and 12003024; and the Strategic Priority Research Program of the Chinese Academy of Science ``Multi-wave band Gravitational Wave Universe'' (No. XDB23040000). X. Z. acknowledges the support from the China Postdoctoral Science Foundation (2019M662233).

\bibliography{main}{}
\bibliographystyle{aasjournal}

\end{document}